\def\aa{A\&A}               %Astronomy & Astrophysics%
\def\aas{A\&AS}             %A & A Supplements%
\def\anj{AJ}                %Astronomical Journal%
\def\apj{ApJ}               %Astrophysical Journal%
\def\mn{MNRAS}              %Monthly Notices of the Royal...%
\def\rev{ARA\&A}            %Annual Review of Astronomy & Astrophysics%
\def\newar{NewAR}           %New Astronomy Reviews

\def\mc{\multicolumn{2}{c}{---}}

\documentstyle{aa}  

\begin{document}

\title{A new sample of large angular size radio galaxies}
\subtitle{II. The optical data}

\author{L. Lara\inst{1}\thanks{Visiting Astronomer, German-Spanish
Astronomical Center, Calar Alto, operated by the Max-Planck-Institut
f\"ur Astronomie jointly with the Spanish National Commission for
Astronomy} \and 
I. M\'arquez\inst{1\star} \and
W.D. Cotton\inst{2} \and
L. Feretti\inst{3} \and
G. Giovannini\inst{3,4} \and
J.M. Marcaide\inst{5} \and
T. Venturi\inst{3}}

\offprints{L. Lara \email{lucas@iaa.csic.es}}

\institute{Instituto de Astrof\'{\i}sica de Andaluc\'{\i}a (CSIC),
Apdo. 3004, 18080 Granada (Spain)
\and 
National Radio Astronomy Observatory, 520 Edgemont Road, Charlottesville, 
VA 22903-2475 (USA)
\and
Istituto di Radioastronomia (CNR), via P. Gobetti 101, 40129 Bologna (Italy)
\and
Dipartamento di Fisica, Universit\'a di Bologna, via B. Pichat 6/2,
40127 Bologna (Italy)
\and
Departimento de Astronom\'{\i}a, Universitat de Val\`encia, 46100 Burjassot - 
Spain
} 
\date{Received / Accepted}

\authorrunning{Lara et al.}
\titlerunning{A new sample of large angular size radio galaxies. II}

\abstract{
We constructed and presented in the first paper of this series a new
sample of 84 large angular size radio galaxies by selecting from the
NRAO VLA Sky Survey objects with angular size $\ge 4\arcmin$,
declination above $+60^{\circ}$ and total flux density at 1.4 GHz $\ge
100$ mJy.  In this paper we present optical spectra and images of
those galaxies associated with the radio emission for which no
redshift was known prior to our observations.  Optical counterparts
have been identified for all (but one) members of the sample. After
our observations, a reliable spectroscopic redshift is available for
67 objects (80\%) from the sample. This paper, second of a series of
three, contributes to increase the number of well defined samples of
radio galaxies with ample spectroscopic information.
}

\maketitle

\section{Introduction}

Lara et al. (\cite{paperI}), hereafter Paper I, presented a new sample
of large angular size radio galaxies selected from the
NRAO\footnote{National Radio Astronomy Observatory} VLA\footnote{Very
Large Array, operated by the NRAO} Sky Survey (NVSS; Condon et
al. \cite{nvss}). The sample, covering an sky area of $\pi$
steradians, was constructed under the following selection criteria:
declination above $+60^{\circ}$, total flux density at 1.4 GHz greater
than 100 mJy and angular size larger than 4$\arcmin$. 122 radio
sources were pre-selected and observed with the VLA with higher
angular resolution at 1.4 and 4.9 GHz for confirmation, and a total of
84 radio galaxies were selected for the final sample (see Table 2 in
Paper I). In Paper I we have shown that, within the selection
criteria, our sample is homogeneous and suitable for statistical
studies. Moreover, our sample represents a substantial increase in the
number of known giant radio galaxies (GRGs; defined as those with a
projected linear size\footnote{We assume that H$_{0}=50$km s$^{-1}$
Mpc$^{-1}$ and q$_{0}=0.5$} $\ge$ 1 Mpc), with 22 new objects
belonging to this class. This sample, together with other new samples
being currently studied by other groups (Schoenmakers et al. \cite{arno1},
\cite{arno2}; Machalski et al. \cite{machalski}) raises the number of known 
GRGs to above one hundred.

The identification of the optical counterpart of the radio core
emission, and the determination of its redshift is a necessary step
for the study of any sample of radio galaxies.  In our case, many of
the members of the sample were poorly known, both at radio and optical
wavelengths. First, a reliable optical identification of the radio
galaxies had not been possible previously to our work since only low
angular resolution radio data existed and, consequently, only a poor
determination of the radio core position was possible. The radio maps
of the sample members presented in Paper I allowed us to obtain
accurate positions of the core components (see Table~\ref{sample})
and, in most cases, to attempt successfully the identification of the
associated galaxy on the Digital Sky Survey (DSS).  Second, only 35\%
of the galaxies in the sample had measured redshifts. Thus, it was
necessary to tackle the task of measuring the redshift of all the
remaining galaxies in order to obtain useful physical parameters for a
subsequent study.

We present here (Paper II) images obtained with the 2.2m telescope in
Calar Alto of the optical counterparts of the sample galaxies for which
no redshift was available in the literature at the time of the
observations.  We have made long-slit spectroscopy of these objects
and determined the redshift for 46 (6 uncertain) of them.  In a
forthcoming third paper of this series, we will present a global
analysis of the sample properties.

\section{Optical imaging and spectroscopy}

We made optical observations of 57 radio galaxies from the sample
pointing the 2.2m telescope in Calar Alto (Spain) to the position of
the radio core component. The log of the observing runs is given in
Table~\ref{obs}. We used the Calar Alto Faint Object Spectrograph
(CAFOS) equipped with a SITe-1d $2048\times2048$ CCD. CAFOS allows
direct imaging and spectroscopy, with a spatial scale of $0''.53$ per
pixel with this CCD.  To obtain the spectra, we used a medium
resolution grism (200\AA/mm), sensitive to the wavelength range of
4000 to 8500 \AA~, that provides a spectral resolution of
4.47\AA/pixel.

The standard procedure followed during the observations was to take
first one 300s exposure image in the Johnson R-band in order to
identify the optical host of the radio galaxy core component, whose
position had already been accurately determined from VLA radio
observations (Paper I) and registered to the DSS plates. Once the
galaxy coincident with the radio core was located in the image, we
obtained a spectrum with the CAFOS long-slit configuration, taking two
equal exposures (usually 900s, or 1200s for weaker objects) to be
added up in order to reject the cosmic rays. We used a $2''$ wide slit
placed north-south, except in J1220+636, J1251+756, J0819+756 and
J1251+787, for which a $-45^{\circ}$ slit rotation was applied in
order to avoid contamination from nearby bright stars.  Some objects
required larger exposures in the R-band in order to detect the
associated optical galaxy, and in some of the weaker galaxies,
although detected, we did not attempt to obtain an spectrum.

The data reduction and calibration were performed following standard
procedures with the IRAF\footnote{IRAF is the Image Reduction and
Analysis Facility made available to the astronomical community by the
National Optical Astronomy Observatories, which are operated by the
Association of Universities for Research in Astronomy (AURA), Inc.,
under contract with the U.S.  National Science Foundation.}  software,
involving dark and flat field corrections.  Wavelength calibration was
carried out using exposures of mercury-helium-rubidium arc lamps taken
just before or after the target exposure.  No flux calibration of the
spectra was attempted since the atmospheric conditions were far from
photometric during most of the observing nights. In fact, some of the 
resulting spectra have quite a low signal to noise ratio due to bad weather 
conditions (e.g. J0757+826 or J0828+632). Examining the
intensity profile along the slit on each spectrum, we determined the
most adequate region for the determination and subtraction of the sky
contribution, and the region centered on the target galaxy which we
summed in order to create the corresponding 1-dimension spectrum.

Images of the optical counterpart of the observed radio galaxies and
their spectra are shown in Fig.~\ref{fig1}.  The average seeing (FWHM)
determined for each night using foreground stars in the different
R-band images is displayed in Table~\ref{obs}, and varies from $2''$
on the night of 30th October 1998 to $1''$ on 9th December 1998.  Two
prominent atmospheric absorption bands at $\lambda = 7604$ \AA~ and
$\lambda = 6882$ \AA, corresponding to the Fraunhofer A and B bands of
molecular oxygen, are marked with vertical lines on the spectra in
Fig.~\ref{fig1}.

\begin{table}[]
\caption[]{Optical observations}
\label{obs}
\begin{tabular}{lccc}
\hline
Nights & Average seeing & $\sigma$ & Code$^a$\\
       &    ($\arcsec$) &          &         \\
\hline
Sep. 26th 1997 & 1.88 & 0.36 & a1 \\
Oct. 30th 1998 & 1.98 & 0.33 & b1 \\
Oct. 31st 1998 & 1.16 & 0.14 & b2 \\
Nov. 1st 1998  & 1.23 & 0.17 & b3 \\
Nov. 2nd 1998  & 1.72 & 0.22 & b4 \\
Dec. 8th 1998  & 1.08 & 0.17 & c1 \\
Dec. 9th 1998  & 1.02 & 0.14 & c2 \\
\hline
\end{tabular}
\parbox{7cm}{
$^a$ Code used in Tab.~\ref{sample} to indicate the epoch of observation 
of each source.}
\end{table}

% ** PONER TAMAGNO FIGURA AL 100% y vspace a 24cm *****
% ** PARA REFEREE, AL 80% Y VSPACE 20CM 
\begin{figure*}
\vspace{24cm}
%\special{psfile=fig1_1.ps hoffset=-50 voffset=-55 hscale=100 vscale=100 angle=0}
%\rule{0.4pt}{4cm}% line thickness, height of picture

\caption{Optical images of the galaxies identified with the radio sources of the sample. The white cross (or broken line) helps to identify the galaxy coincident with the position of the radio core. The segment on upper right corresponds to an angular size of $10''$. Below each image we 
display the optical spectrum, obtained with a long slit
of $2''$ width. Flux units are arbitrary and wavelengths are given in \AA. The two vertical solid lines
mark the position of atmospheric absorption bands produced by O$_2$.}

\label{fig1}
\end{figure*}

\begin{figure*}
\vspace{24cm}
\addtocounter{figure}{-1}
%\special{psfile=fig1_2.ps hoffset=-50 voffset=-55 hscale=100 vscale=100 angle=0}
%\rule{0.4pt}{4cm}% line thickness, height of picture
\caption{continued} 
\end{figure*}

\begin{figure*}
\vspace{24cm}
\addtocounter{figure}{-1}
%\special{psfile=fig1_3.ps hoffset=-50 voffset=-55 hscale=100 vscale=100 angle=0}
%\rule{0.4pt}{4cm}% line thickness, height of picture
\caption{continued}
\end{figure*}

\begin{figure*}
\vspace{24cm}
\addtocounter{figure}{-1}
%\special{psfile=fig1_4.ps hoffset=-50 voffset=-55 hscale=100 vscale=100 angle=0}
%\rule{0.4pt}{4cm}% line thickness, height of picture
\caption{continued} 
\end{figure*}

\begin{figure*}
\vspace{17cm}
\addtocounter{figure}{-1}
%\special{psfile=fig1_5.ps hoffset=-50 voffset=-255 hscale=100 vscale=100 angle=0}
%\rule{0.4pt}{4cm}% line thickness, height of picture
\caption{continued}
\end{figure*}

\section{Notes on singular sources}

In this section we make a brief description of those observed
galaxies which show some type of remarkable properties in the optical
regime (luminosity, spectrum or nearby environment), when
compared with the rest of the sample:

{\bf J0342+636}: A spheroidal galaxy with a prominent broad H$\alpha$
emission line. A weaker broad H$\beta$ line and [OIII] narrow emission
lines are also detected. J0342+636 presents a close companion at an
angular distance of $6.8''$ in position angle (P.A.)  of
$-53^{\circ}$, which corresponds to a linear projected distance of
20.5 kpc (if both galaxies are assumed to be at the same redshift).

{\bf J0455+603}: At the time of the optical observations we had not
correctly identified the core component in the radio structure, and
consequently did not find the associated galaxy after two exposures of
300 s each. The core component of J0455+603 was misidentified with a
bright and compact feature in our 1.4 GHz VLA map (see Paper
I). However, we have now identified the core with a weaker component
with flat spectrum at $16''$ in NW direction from the previously 
assumed core, which is
coincident with the galaxy marked in Fig.~\ref{fig1}. Coordinates in
Table~\ref{sample} have been corrected with respect to those listed in
Paper I.

{\bf J0502+670}: The host galaxy of this Fanaroff-Riley type I radio
source (FR I; Fanaroff \& Riley \cite{fanaroff}) is a bright
elliptical at z=0.085 with an spectrum characterized by the presence
of stellar absorption features. It has a close companion separated by
$12''$ at P.A. of 29$^{\circ}$, which corresponds to 25.7 kpc if
assumed to be at the same redshift. There is a foreground star
($m_{R}\sim 13$) between the two galaxies.

{\bf J0525+718}: A bright elliptical galaxy is the source of this low
power radio galaxy. There are two nearby galaxies at $7.6''$ in
P.A. $+2^{\circ}$ and at $12''$ in P.A. $-21^{\circ}$, which if assumed to be at the same redshift of J0525+718 correspond to projected distances of 26 and 
41 kpc, respectively.

{\bf J0750+656}: The most distant radio source in our sample, at
z=0.747. It is optically identified with a 16.4 magnitude quasar type object. 

\begin{figure}
\vspace{4cm}
\includegraphics{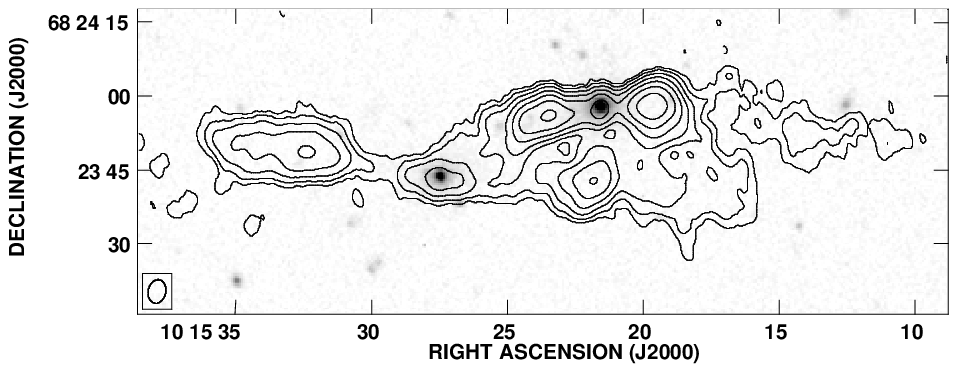}
%\rule{0.4pt}{4cm}% line thickness, height of picture
\caption{Superposition of a radio image of J1015+683 at 4.9 GHz (Paper I) with
the Johnson R-band filter optical image of the same field.}  
\label{fig2}
\end{figure}

{\bf J1015+683}: Two galaxies in the field can be identified with
components in the complex radio structure of this source (see
Figure~\ref{fig2}), which confirms that J1015+683 results from the
superposition of two distinct radio galaxies with their main axes in a
similar (projected) direction. The angular distance of the two
galaxies is $35.4''$, which at the redshift of the northern one
(z=0.199) corresponds to 149 kpc. The classification of this radio
galaxy as a giant (Paper I) must be postponed until a redshift 
determination of the southern galaxy. 

{\bf J1137+613:} This radio galaxy, of Fanaroff-Riley type II (FR II),
originates from an spheroidal galaxy. Its spectrum presents 
high signal to noise ratio emission lines (H$\beta$, [OIII], [OI], H$\alpha$, [NII], [SII]). 

{\bf J1211+743}: The radio emission originates from a bright
elliptical galaxy. It has a nearby companion at $7.5''$ in
P.A. $+51^{\circ}$, which corresponds to 19.5 kpc if assumed to be at
the same redshift of J1211+743. There are not emission features in the
spectrum, which is dominated by the continuum and some stellar absorption
lines.

{\bf J1800+717}: It is the only object in our sample for which no hint
of the optical counterpart of the radio emission is found. The cross
in Fig.~\ref{fig1} represents the position of the radio core of this
FR II radio galaxy. We made three 300s exposures on the field
centered at this position but nothing was detected.  The R-band
extinction is only of 0.1 magnitudes in this region (Schlegel et
al. \cite{schlegel}), so the associated galaxy must either be very
distant or intrinsically weak. A limit to its apparent magnitude (m$_R >
19.5$) can be estimated from the magnitude of a weak object located
$13''$ northward from the radio core position.

{\bf J1845+818}: It is identified with an elliptical galaxy.  At a
distance of $4.4''$ in P.A. $+76^{\circ}$ from the galaxy center,
there is a small, possibly elongated feature. It is coincident with
the direction of the radio jet, although if it is related to the jet
is not clear. Most probably, it is a small galaxy close to the
radio emitting elliptical galaxy, at a projected distance of 12.5 kpc
if assumed to be at the same redshift. The galaxy spectrum presents
only stellar absorption lines.

{\bf J2059+627}: We obtained only a bad quality optical image of this
field, although sufficient to identify a very weak feature associated
with the radio core. Two 1200s long-slit exposures revealed a weak
continuum with prominent [OIII] emission lines.

{\bf J2114+820}: A peculiar low power radio galaxy of FR I type
associated with an elliptical galaxy which shows prominent broad
H$\alpha$, H$\beta$ and H$\gamma$ emission lines, and narrow [OIII]
lines (Lara et al. \cite{lara2}). The galaxy has a bright compact core
associated to the nuclear activity.

{\bf J2138+831}: The origin of this radio source is a bright
elliptical galaxy residing in the cluster Abell 2387. Its spectrum 
shows no emission lines, being characterized by stellar absorption lines.

{\bf J2157+664}: We made three 300s exposures on this galaxy which
revealed an extended elliptical galaxy with an apparent double core in
east-west direction. The radio core is coincident with the western
feature. No emission lines are found in the spectrum. The galaxy, with
a galactic latitude of $+9.2^{\circ}$, lies in a crowded field, with
possibly smaller galaxies around, and a bright foreground star located
nearby southwards.

{\bf J2247+633}: We made two 300s exposures on the field centered at
the radio core position, and found a very weak feature coincident with
the radio core. The galactic latitude of this galaxy is only of
$+3.7^{\circ}$. We did not attempt to obtain the spectrum of this
galaxy.

{\bf J2250+729}: We made three 300s exposures on the field centered at
the radio galaxy core. A very weak feature associated with the radio
core was found, close to a bright star with apparent m$_R = 10.1$. The
galactic latitude of this galaxy is $+12.1^{\circ}$. We did not
attempt to obtain the spectrum of this galaxy.

{\bf J2340+621}: We made two 300s exposures and tentatively found the
optical counterpart of this radio galaxy in a crowded star field. The
galactic latitude is only $+0.4^{\circ}$ and the R-band extinction in
this region is of 4.0 magnitudes (Schlegel et al. \cite{schlegel}). We
did not attempt to obtain the optical spectrum of this galaxy.

\section{Results}

\begin{table*}[t]
\caption[]{Sample of large angular size radio galaxies from the NVSS survey}
\label{sample}
%\begin{footnotesize}
\begin{scriptsize}
\begin{tabular}{lcc r@{.}l clll r@{.}l r@{.}l l}
~~~Name & R.A.(J2000.0)&Dec.(J2000.0)  & \multicolumn{2}{c}{b} &Epoch & T$_{exp}$ & ~~z & Notes & \multicolumn{2}{c}{m$_R$} & \multicolumn{2}{c}{$r_e$}  &  Radio \\ 
        &($^h~~^m~~^s$)&($^{\circ}$~~$\arcmin$~~$\arcsec$)& \multicolumn{2}{c}{($^{\circ}$)} & & ~~(s) & &   & \multicolumn{2}{c}{(mag)} & \multicolumn{2}{c}{(kpc)} & Type \\
\hline	 
\object{J0109+731} & 01 09 44.265 & 73 11 57.17 & 10&4 &    &        & 0.181$^a$      & SEL & \mc      & \mc  & II    \\
\object{J0153+712} & 01 53 25.786 & 71 15 06.53 &  9&0 &    &        & 0.022$^a$      & SEL & 10&4     & \mc  & I     \\
\object{J0317+769} & 03 17 54.061 & 76 58 37.82 & 16&5 & b1 & 2x900  & 0.094          & AbL & 14&7     & 37&0 & I     \\
\object{J0318+684} & 03 18 19.026 & 68 29 32.08 &  9&4 &    &        & 0.090$^b$      & WEL & 17&5     & \mc  & II    \\
\object{J0342+636} & 03 42 10.148 & 63 39 33.73 &  6&8 & a1 & 2x900  & 0.128          & SEL & 17&9     & \mc  & II    \\
\object{J0430+773} & 04 30 49.490 & 77 22 58.44 & 19&5 & b1 & 2x1200 & 0.215          & WEL & 18&5     & 10&4 & II    \\
\object{J0455+603} & 04 55 45.847 & 60 23 48.94 & 10&6 & c1 &        & ---            & --- & \mc      & \mc  & I     \\
\object{J0502+670} & 05 02 54.732 & 67 02 30.15 & 15&2 & b2 & 2x1200 & 0.085          & AbL & 13&9     & 28&9 & I     \\
\object{J0508+609} & 05 08 27.258 & 60 56 27.48 & 12&2 & b2 & 2x900  & 0.071          & WEL & 14&6     & 34&6 & I     \\
\object{J0519+702} & 05 19 17.132 & 70 13 48.68 & 18&1 & b1 & 2x900  & 0.144          & AbL & 16&1     & 33&1 & I     \\
\object{J0525+718} & 05 25 27.094 & 71 52 39.25 & 19&4 & b2 & 2x900  & 0.150          & AbL & 15&9     & 26&8 & I     \\
\object{J0531+677} & 05 31 25.925 & 67 43 50.23 & 17&9 & a1 & 2x900  & 0.017          & AbL &  8&8     &  3&0 & I     \\
\object{J0546+633} & 05 46 24.622 & 63 21 32.50 & 17&2 &    &        & 0.049$^a$      & ??  & 10&9     & \mc  & I     \\
\object{J0559+607} & 05 59 38.690 & 60 44 00.96 & 17&5 & b2 & 2x900  & 0.042          & AbL & 12&6     & 14&3 & I     \\
\object{J0607+612} & 06 07 34.919 & 61 14 43.52 & 18&6 & b3 & 2x1200 & 0.227          & WEL & \mc      & \mc  & I/II  \\
\object{J0624+630} & 06 24 29.063 & 63 04 02.50 & 21&1 & b3 & 2x900  & 0.085          & AbL & 14&5     & 10&4 & I     \\
\object{J0633+721} & 06 33 40.842 & 72 09 24.92 & 24&4 &    &        & 0.090$^b$      & AbL & 15&0     & \mc  & I/II  \\
\object{J0654+733} & 06 54 26.525 & 73 19 50.36 & 26&1 &    &        & 0.115$^b$      & SEL & 15&2     & \mc  & II    \\
\object{J0750+656} & 07 50 34.425 & 65 41 25.50 & 30&6 & b1 & 2x900  & 0.747          & SEL & 16&4     & \mc  & II-QSS\\
\object{J0757+826} & 07 57 35.172 & 82 39 40.86 & 29&0 & b4 & 3x1200 & 0.087? (0.06)  & AbL & 12&0     &  9&6 & I     \\
\object{J0803+669} & 08 03 45.829 & 66 56 11.39 & 31&9 & b3 & 2x1200 & 0.247? (0.37)  & AbL & 19&3     & \mc  & II    \\
\object{J0807+740} & 08 07 10.070 & 74 00 41.58 & 31&2 &    &        & 0.120$^b$      & SEL & 15&8     & \mc  & I     \\
\object{J0819+756} & 08 19 50.504 & 75 38 39.53 & 31&7 & c2 & 1x900  & 0.232$^b$      & SEL & 17&7     & 24&8 & II    \\
\object{J0825+693} & 08 25 59.770 & 69 20 38.59 & 33&5 &    &        & 0.538$^a$      & SEL & \mc      & \mc  & II    \\
\object{J0828+632} & 08 28 56.363 & 63 13 45.05 & 34&9 & b4 & 2x1200 & ---~~~~~~(0.09)& --- & 13&7     &(10&7)& I/II  \\
\object{J0856+663} & 08 56 16.260 & 66 21 26.50 & 37&1 & b3 & 2x1200 & 0.489          & WEL & 19&7     & \mc  & II    \\
\object{J0926+653} & 09 26 00.822 & 65 19 22.88 & 40&3 & b2 & 2x900  & 0.140          & AbL & 15&0     & 42&3 & I     \\
\object{J0926+610} & 09 26 53.408 & 61 00 24.87 & 42&0 & b2 & 1x900  & 0.243          & SEL & 17&2     & 25&8 & II    \\
\object{J0939+740} & 09 39 46.833 & 74 05 30.78 & 37&0 &    &        & 0.122$^b$      & AbL & 14&8     & \mc  & I     \\
\object{J0949+732} & 09 49 46.157 & 73 14 23.82 & 38&1 &    &        & 0.058$^a$      & SEL & 12&8     & \mc  & II    \\
\object{J1015+683} & 10 15 21.620 & 68 23 58.24 & 42&7 & c1 & 2x900  & 0.199          & AbL & 15&1     & 31&6 & ?     \\
\object{J1036+677} & 10 36 41.237 & 67 47 53.44 & 44&6 & c1 &        & ---            & --- & \mc      & \mc  & II    \\
\object{J1124+749} & 11 24 47.045 & 74 55 45.31 & 40&9 & c1 & 1x900  & 0.052          & WEL & 10&7     & 34&9 & I     \\
\object{J1137+613} & 11 37 21.289 & 61 20 01.88 & 53&6 & c1 & 2x900  & 0.111          & SEL & 16&2     & 18&8 & II    \\
\object{J1211+743} & 12 11 58.710 & 74 19 04.12 & 42&5 & c1 & 2x900  & 0.107          & AbL & 12&6     & 49&1 & I/II  \\
\object{J1216+674} & 12 16 37.239 & 67 24 41.97 & 49&4 & c2 & 2x900  & 0.362          & SEL & 18&7     & 41&0 & II    \\
\object{J1220+636} & 12 20 36.477 & 63 41 43.82 & 53&1 & c2 & 2x900  & 0.188          & AbL & 15&7     & \mc  & II    \\
\object{J1247+673} & 12 47 33.319 & 67 23 16.34 & 49&7 &    &        & 0.107$^a$      & AbL & 14&7     & \mc  & II    \\
\object{J1251+756} & 12 51 05.977 & 75 37 38.94 & 41&5 & c2 & 2x900  & 0.197          & WEL & 16&4     &  6&9 & II    \\
\object{J1251+787} & 12 51 23.839 & 78 42 36.29 & 38&4 & c2 & 900+700& 0.045          & AbL & 10&4     & 17&1 & I     \\
\object{J1313+696} & 13 13 58.878 & 69 37 18.74 & 47&4 &    &        & 0.106$^a$      & SEL & 16&0     & \mc  & II    \\
\object{J1410+633} & 14 10 30.609 & 63 19 00.55 & 51&6 & c1 & 2x900  & 0.158          & AbL & 17&2     & 13&4 & II    \\
\hline 
\multicolumn{14}{l}{
\parbox{16cm}{
$^a$ Redshift taken from the NASA Extragalactic Database\\
$^b$ Redshift from Schoenmakers et al (\cite{arnotesis})
}}
\end{tabular}
%\end{footnotesize}
\end{scriptsize}
\end{table*}

\begin{table*}[t]
\addtocounter{table}{-1}
\caption[]{continued}
\begin{scriptsize}
\begin{tabular}{lcc r@{.}l clll r@{.}l r@{.}l l}
~~~Name & R.A.(J2000.0)&Dec.(J2000.0)  & \multicolumn{2}{c}{b} &Epoch & T$_{exp}$ & ~~z & Notes & \multicolumn{2}{c}{m$_R$} & \multicolumn{2}{c}{$r_e$}  &  Radio \\ 
        &($^h~~^m~~^s$)&($^{\circ}$~~$\arcmin$~~$\arcsec$)& \multicolumn{2}{c}{($^{\circ}$)} & & ~~(s) & &   & \multicolumn{2}{c}{(mag)} & \multicolumn{2}{c}{(kpc)} & Type \\
\hline	 
\object{J1504+689} & 15 04 12.781 & 68 56 12.75 & 43&9 &    &        & 0.318$^a$      & SEL & 16&2     &  \mc & II-QSS\\
\object{J1523+636} & 15 23 45.900 & 63 39 23.78 & 46&0 &    &        & 0.204$^a$      & SEL & 16&0     &  \mc & II    \\
\object{J1530+824} & 15 30 56.110 & 82 27 21.02 & 32&8 &    &        & 0.021$^a$      & ??  &  9&1     &  \mc & I     \\
\object{J1536+843} & 15 36 57.335 & 84 23 10.42 & 31&3 &    &        & 0.201$^b$      & WEL & 18&2     &  \mc & II    \\
\object{J1557+706} & 15 57 30.190 & 70 41 20.79 & 39&3 &    &        & 0.026$^a$      & AbL &  8&2     &  \mc & I     \\
\object{J1632+825} & 16 32 31.630 & 82 32 16.28 & 31&2 &    &        & 0.023$^a$      & SEL &  8&4     &  \mc & I     \\
\object{J1650+815} & 16 50 58.686 & 81 34 28.11 & 31&1 &    &        & 0.038$^a$      & ??  & 11&3     &  \mc & I     \\
\object{J1732+714} & 17 32 33.001 & 71 24 10.50 & 31&9 &    &        & 0.059$^a$      & ??  & 12&0     &  \mc & I     \\
\object{J1733+707} & 17 33 12.525 & 70 46 30.36 & 31&9 &    &        & 0.041$^a$      & ??  & 11&4     &  \mc & I     \\
\object{J1743+712} & 17 43 17.681 & 71 12 53.98 & 31&0 &    &        & ---~~~~~~(0.25)& --- & 17&8     &  \mc & II    \\
\object{J1745+712} & 17 45 43.573 & 71 15 48.55 & 30&8 &    &        & 0.216$^a$      & SEL & 18&4     &  \mc & II    \\
\object{J1751+680} & 17 51 19.629 & 68 04 43.05 & 30&6 & b4 & 2x900  & 0.079          & AbL & 12&4     & 43&6 & I     \\
\object{J1754+626} & 17 54 50.310 & 62 38 41.96 & 30&3 &    &        & 0.028$^a$      & WEL & 10&3     &  \mc & I     \\
\object{J1800+717} & 18 00 42.622 & 71 44 41.99 & 29&6 & b4 &        & ---~~~~($\geq$0.40)& --- & $\geq$19&5 &  \mc & II    \\
\object{J1835+665} & 18 35 07.338 & 66 35 00.02 & 26&3 & b3 & 2x1200 & 0.354? (0.34)  & AbL & 19&0     & 14&6 & II    \\
\object{J1835+620} & 18 35 10.405 & 62 04 07.42 & 25&6 & b1 & 2x1200 & 0.518          & SEL &  \mc     &  \mc & II    \\
\object{J1844+653} & 18 44 07.443 & 65 22 03.07 & 25&2 & b1 & 2x1200 & 0.197          & AbL & 17&1     &  \mc & II    \\
\object{J1845+818} & 18 45 15.836 & 81 49 30.98 & 27&0 & b1 & 2x900  & 0.119          & AbL & 14&3     & 30&6 & II    \\
\object{J1847+707} & 18 47 34.912 & 70 44 00.64 & 25&8 & a1 & 500+900& 0.043          & AbL & 10&6     & 13&4 & I     \\
\object{J1850+645} & 18 50 45.871 & 64 30 24.68 & 24&4 & a1 & 2x900  & 0.080          & AbL & 12&6     & 21&6 & I     \\
\object{J1853+800} & 18 53 52.077 & 80 02 50.46 & 26&6 &    &        & 0.214$^a$      & ??  & 17&8     &  \mc & II    \\
\object{J1918+742} & 19 18 34.885 & 74 15 05.05 & 24&2 & b2 & 2x1200 & 0.194          & WEL & 17&5     &  8&2 & II    \\
\object{J1951+706} & 19 51 40.825 & 70 37 39.99 & 20&7 & b3 & 2x1200 & 0.550          & SEL &  \mc     &  \mc & II    \\
\object{J2016+608} & 20 16 18.630 & 60 53 57.49 & 14&0 & c1 & 2x900  & 0.121          & SEL & 18&2     & 10&6 & II    \\
\object{J2035+680} & 20 35 16.549 & 68 05 41.60 & 16&1 & b2 & 2x1200 & 0.133          & WEL & 18&0     &  9&9 & I     \\
\object{J2042+751} & 20 42 37.180 & 75 08 02.52 & 19&5 &    &        & 0.104$^a$      & SEL & 14&7     &  \mc & II-QSS\\
\object{J2059+627} & 20 59 09.560 & 62 47 44.11 & 11&0 & c2 & 2x1200 & 0.267          & SEL &  \mc     &  \mc & II?   \\
\object{J2103+649} & 21 03 13.868 & 64 56 55.26 & 12&0 & c2 & 2x1200 & 0.215          & AbL &  \mc     &  \mc & II    \\
\object{J2111+630} & 21 11 20.268 & 63 00 06.17 & 10&1 & c1 & 2x900  & ---            & --- &  \mc     &  \mc & II    \\
\object{J2114+820} & 21 14 01.179 & 82 04 48.28 & 22&3 & a1 & 2x900  & 0.085          & SEL & 13&5     &  \mc & I     \\
\object{J2128+603} & 21 28 02.634 & 60 21 07.96 &  6&8 & b2 & 2x1200 & 0.072?         & AbL &  \mc     &  \mc & II    \\
\object{J2138+831} & 21 38 42.266 & 83 06 49.21 & 22&4 & b2 & 2x900  & 0.135          & WEL & 14&5     & 36&9 & I/II  \\
\object{J2145+819} & 21 45 29.887 & 81 54 54.22 & 21&4 &    &        & 0.146$^b$      & SEL & 17&1     &  \mc & II    \\
\object{J2157+664} & 21 57 02.572 & 66 26 10.24 &  9&2 & b3 & 2x900  & 0.057?         & AbL &  \mc     &  \mc & I/II  \\
\object{J2204+783} & 22 04 09.225 & 78 22 46.92 & 18&2 & a1 & 2x900  & 0.115          & AbL & 15&1     & 15&5 & II    \\
\object{J2209+727} & 22 09 33.780 & 72 45 58.36 & 13&6 & c2 & 2x1200 & 0.201          & WEL &  \mc     &  \mc & II    \\
\object{J2242+622} & 22 42 32.133 & 62 12 17.53 &  3&0 & c1 & 2x1200 & 0.188?         & AbL &  \mc     &  \mc & II    \\
\object{J2247+633} & 22 47 29.714 & 63 21 15.55 &  3&7 & c2 &        & ---            & --- &  \mc     &  \mc & I     \\
\object{J2250+729} & 22 50 43.621 & 72 56 16.19 & 12&1 & c1 &        & ---            & --- &  \mc     &  \mc & II    \\
\object{J2255+645} & 22 55 29.943 & 64 30 06.86 &  4&4 & c2 &        & ---            & --- &  \mc     &  \mc & II    \\
\object{J2307+640} & 23 07 58.533 & 64 01 39.22 &  3&4 & c2 &        & ---            & --- &  \mc     &  \mc & II    \\
\object{J2340+621} & 23 40 56.435 & 62 10 45.09 &  0&4 & b3 &        & ---            & --- &  \mc     &  \mc & I     \\
\hline
\multicolumn{14}{l}{
\parbox{16cm}{
$^a$ Redshift taken from the NASA Extragalactic Database\\
$^b$ Redshift from Schoenmakers (\cite{arnotesis})
}} 
\end{tabular}
\end{scriptsize}
\end{table*}

In Table~\ref{sample} we present the main results derived from the 
observations reported in this paper:\\
{\em Column 1}: IAU source name at epoch J2000.\\
{\em Column 2 - 3}: right ascension and declination (J2000).\\
{\em Column 4}: galactic latitude.\\ 
{\em Column 5}: code defined in Table~\ref{obs} indicating the epoch of observation.\\
{\em Column 6}: integration time for the spectroscopic observations.\\ 
{\em Column 7}: redshift derived from our observations or from the literature.\\ 
{\em Column 8}: notes about the optical spectrum: SEL - strong
emission lines present; WEL - weak emission lines; AbL - absorption lines
only.\\
{\em Column 9}: apparent photographic magnitudes derived from the
red band Palomar survey (DSS R-band). Magnitudes of faint galaxies
(above the 16th magnitude) have been obtained from the USNO-A2.0
catalogue (Monet et al. \cite{monet}), and compared with the APM
catalogue ({\tt http://www.ast.cam.ac.uk/\~{}apmcat}). In general, we
found good agreement between these two catalogues, but we adopted the
mean value in the few cases where the discrepancy was larger than 0.4
magnitudes. When USNO-A2.0 did not provide information on a requested
galaxy, we adopted the APM magnitude. The magnitudes of brighter
galaxies (up to the 16th magnitude) have been determined directly from
the DSS using the available photometric calibration (Doggett et
al. \cite{doggett}). However, we note that the apparent magnitudes of
the brighter extended galaxies are subject of large errors (sometimes
larger than 3 magnitudes) probably due to the non-linear behavior of the
photographic plates.\\
{\em Column 10}: effective radius of the galaxies derived by fitting a
$r^{1/4}$ profile to the brightness distribution. Although the low
sensitivity or quality of some galaxy images prevented us from
obtaining reasonable fits for them, in all possible cases the profile
was consistent with that typical of elliptical galaxies.\\
{\em Column 11}: type of radio structure: I and II stand for FR I and FR II type radio galaxies, respectively. Distinction between the two classes is based solely on morphological aspects. Those radio galaxies with a difficult 
distinction between the FR I or FR II types are labelled as I/II.\\

\begin{figure}
\vspace{6cm}
\includegraphics{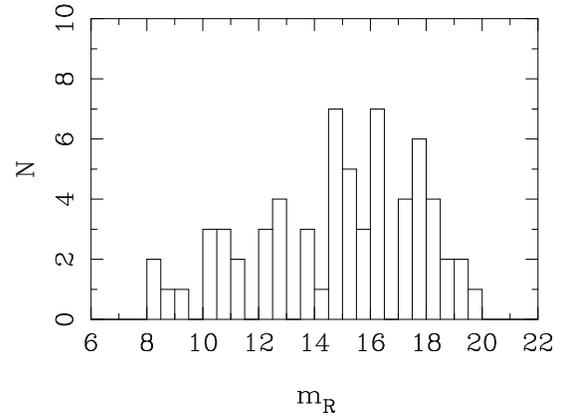}
%\rule{0.4pt}{4cm}% line thickness, height of picture
\caption{Histogram of the apparent R-band magnitudes of the galaxies in 
our sample. The bin size is 0.5 magnitudes.}  
\label{histo_mag}
\end{figure}

We plot in Fig.~\ref{histo_mag} a histogram of the apparent R-band
magnitude of the galaxies in the sample.  The magnitude distribution
is very extended, ranging from 8.2 up to magnitude 19.7 with a peak
lying between the 15th and 16th magnitudes.  The distribution
presents a tail towards brighter objects induced by the poor calibration 
of photographic plates due to their non-linear behavior.
At the other extreme the distribution is influenced by the
limiting sensitivity of the DSS ($m_R\leq 20$).

\begin{figure}
\vspace{7cm}
\includegraphics{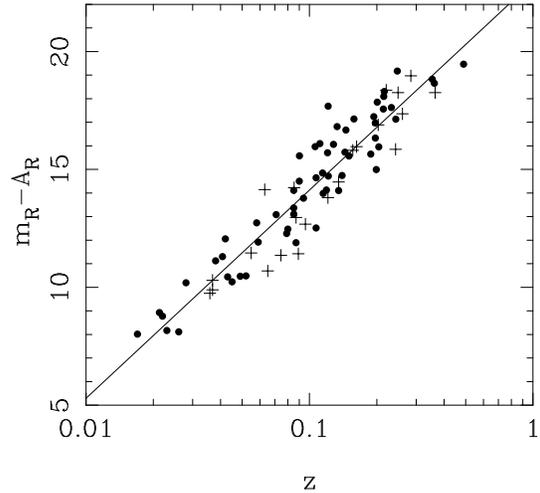}
%\rule{0.4pt}{4cm}% line thickness, height of picture
\caption{De-reddened apparent R-band magnitudes of the members of our sample (dots) and Schoenmakers' et al. sample (crosses), plotted against their redshifts. The solid line represents a linear fit to the data.}  
\label{mag_logz}
\end{figure}

In Fig.~\ref{mag_logz} we represent the apparent de-reddened R-band
magnitude as a function of redshift for the sample members, excepting the three quasar type objects in the sample. We have also added in this plot the radio galaxies (excepting quasars) from the sample constructed by Schoenmakers et al. (\cite{arno2}), who provide a photometry consistent with ours. We find a relation of
the form (least squares fitting)
\begin{equation}
m_R - A(R) = (8.83\pm0.35)\log{z} + (22.96\pm 0.37) 
\label{relation}
\end{equation}
with a correlation coefficient $r=0.942$ and rms$= 1.01$ mag. It allows us to estimate, with a typical error of
$\sim 30$\%, the redshift of those galaxies with known magnitudes
for which an spectroscopic determination was not possible (values
shown in parenthesis in Table~\ref{sample}). For those galaxies with
an uncertain spectroscopic redshift, a ``photometric'' estimation is also
provided when possible (shown in parenthesis).

The fact that such a tight correlation between apparent magnitude and
redshift exists is an indication that the host galaxies of the sample
members have quite similar properties and that a possible orientation
dependent beaming of the core emission does not play a significant
role as expected for AGNs with their main axis oriented close to the
plane of the sky, contrary to favorable oriented objects like
blazars.

\section{Conclusions}

We have made optical observations pointing at the core
position of 57 radio galaxies from our sample and in all cases, except
one, we have detected the optical galaxy responsible of the radio
emission.  Even the galaxy J2340+621, at the very low galactic
latitude of $+0.4^{\circ}$ has been tentatively detected. The observed
galaxies have profiles consistent with being elliptical. We took
spectra of 48 galaxies, of which the redshift could be determined with
high confidence for 40 of them and with uncertainty for 6
galaxies. For only 2 galaxies, J0828+632 and J2111+630, we were unable
to determine the redshift. However, a ``photometric'' redshift can be
estimated from a correlation derived between apparent magnitudes and
spectroscopic redshift of the sample members.  

Regarding the properties of the optical spectrum, 36\% of the galaxies
present spectra with prominent emission lines, 19\% of the galaxies
show weak emission lines with respect to the continuum emission, and
45\% of the galaxies have spectra characterized by stellar absorption
lines only.

The fact that our sample is designed to select radio galaxies with
their main axes oriented on the plane of the sky is consistent with
the spectroscopic results, since only 3 galaxies (J0342+636, J0750+656
and J2114+820) out of 48 observed present broad emission lines, proper
of objects oriented towards the observer according to current
unification schemes (e.g. Antonucci \cite{antonucci}).

In a forthcoming paper of this series, we will discuss the possible
relations between the different parameters derived from the radio and
optical observations presented in this paper and in Paper I, and the
possible dependencies of these parameters with the type of radio
structure.

\begin{acknowledgements}

We thank F. Govoni and I. Gonz\'alez-Serrano for helpful discussion. 
This research is supported in part by the Spanish DGICYT (PB97-1164). LF
and GG acknowledges the Italian Ministry for University and Research
(MURST) for financial support under grant Cofin98-02-32. The National
Radio Astronomy Observatory is a facility of the National Science
Foundation operated under cooperative agreement by Associated
Universities, Inc.  This research has made use of the NASA/IPAC
Extragalactic Database (NED) which is operated by the Jet Propulsion
Laboratory, California Institute of Technology, under contract with
the National Aeronautics and Space Administration, and of the Aladin
interactive sky atlas, CDS, Strasbourg, France.

\end{acknowledgements}

\end{document}